\documentclass[amsmath,11pt]{article}

\date{\empty}

\begin{document}

\title{\bf Peculiar motions, accelerated expansion\\ and the cosmological axis}
\author{Christos G. Tsagas\\ {\small Section of Astrophysics,
Astronomy and Mechanics, Department of Physics}\\ {\small Aristotle
University of Thessaloniki, Thessaloniki 54124, Greece}}

\maketitle

\begin{abstract}
Peculiar velocities change the expansion rate of any observer moving relative to the smooth Hubble flow. As a result, observers in a galaxy like our Milky Way can experience accelerated expansion within a globally decelerating universe, even when the drift velocities are small. The effect is local, but the affected scales can be large enough to give the false impression that the whole cosmos has recently entered an accelerating phase. Generally, peculiar velocities are also associated with dipole-like anisotropies, triggered by the fact that they introduce a preferred spatial direction. This implies that observers experiencing locally accelerated expansion, as a result of their own drift motion, may also find that the acceleration is maximised in one direction and minimised in the opposite. We argue that, typically, such a dipole anisotropy should be relatively small and the axis should probably lie fairly close to the one seen in the spectrum of the Cosmic Microwave Background.
\end{abstract}

\section{Introduction}\label{sI}
Realistic observers do not simply follow the smooth universal expansion, but have their own `peculiar' motions as well. The dipolar anisotropy of the Cosmic Microwave Background (CMB), in particular, has been interpreted as the result of our drift flow (at roughly 600~km/sec) relative to the cosmic rest-frame. Peculiar motions are believed to be the result of structure formation and are expected to fade away as we move on to progressively larger lengths. Thus, the current concordance (WMAP5-normalised) cosmological model predicts drift velocities of approximately 100~km/sec on scales larger than $50h^{-1}$~Mpc ($h$ is the Hubble parameter in units of 100~km/sec\,Mpc). Recent observations however seem to indicate bulk flows with significantly larger amplitude and scale~\cite{KA-BKE}-\cite{LTMC}, casting some doubt on the $\Lambda$CDM scenario~\cite{P}. Here, we look into the potential implications of such large-scale drift motions, which are sometimes referred to as `dark flows', for the kinematics of the associated observers.

The first question is whether drifting observers in a perturbed, dust dominated Friedmann-Robertson-Walker (FRW) universe and those following the Hubble expansion could assign different values (and signs) to their respective deceleration parameters. Whether, in particular, it is theoretically possible for a peculiarly moving observer to `experience' accelerated expansion while the universe is actually decelerating. We find that the answer to this question is positive, when the peculiar velocity field adds to the Hubble expansion. In other words, the drifting observer should reside in a region that expands faster than the background universe. Then, around every typical observer in that patch, there can be a section where the deceleration parameter takes negative values and beyond which it becomes positive again. Moreover, even small (relative to the Hubble rate) peculiar velocities can lead to such local acceleration. The principle is fairly simple: two decelerated expansions (in our case the background and the peculiar) can combine to give an accelerating one, as long as the acceleration is `weak' (with $-1<q<0$ -- where $q$ is the deceleration parameter) and not `strong' (with $q<-1$) -- see \S~\ref{ssWSAE} below. Overall, accelerated expansion for a drifting observer does not necessarily imply the same for the universe itself. Peculiar motions can locally mimic the effects of dark energy. Furthermore, the affected scales can be large enough to give the false impression that the whole universe has recently entered an accelerating phase.

We then ask whether contemporary observations of large-scale bulk flows report drift velocities fast enough to support the above scenario. Again, the answer is positive. Although the $\Lambda$CDM model predicts relatively weak peculiar motions, recent independent surveys seem to indicate larger rms bulk velocities on scales close to $100h^{-1}$~Mpc and beyond~\cite{KA-BKE}-\cite{LTMC}. The analysis of~\cite{WFH}, for example, finds velocities greater than $400$~km/sec on the aforementioned lengths. Analogous results were also reached through the complementary methods of~\cite{LTMC}, while faster bulk flows have been reported by~\cite{KA-BKE}. The latter study, in particular, suggests almost constant peculiar velocities of roughly 1000~km/sec extending as far out as 800~Mpc, with no sign of converging beyond this threshold. On smaller lengths (between 30 and 60~Mpc), on the other hand, the work of~\cite{LS} indicated a positive variance in the local Hubble rate of up to 10\%. Using these measurements, we find that drifting observers can experience locally accelerated expansion on scales of several hundred Mpc (perhaps even larger), depending on the density of the region in question.

When all the observers in the drifting patch have peculiar velocities equal to the mean bulk flow of the region, they should see no preferred direction in the distribution of the deceleration parameter within the patch. In reality, however, observers have their own peculiar motions, which are generally different from the average. In principle, this should trigger a dipolar anisotropy in the measured distribution of the deceleration parameter, analogous to the CMB dipole. Now, however, the rest frame is not that of the universal expansion but the one defined by the mean flow of the patch. In other words, there should be a preferred axis in the accelerated domain, with the acceleration maximised in one direction and minimised in the opposite. Nevertheless, when dealing with typical observers -- those with peculiar velocities close to the average one, the aforementioned dipolar anisotropy must be small. Furthermore, the direction of the $q$-dipole should lie fairly close to that of the CMB, since they are both the result of the observers' drift flow. Interestingly, some recent reports seem to indicate that a `cosmological axis', namely a weak dipolar anisotropy aligned close to the CMB dipole, might actually reside within the supernovae data~\cite{KL}.

We begin by outlining the basic features of the peculiar kinematics in linearly perturbed, dust-dominated FRW universes. Our first aim is to show that drifting observers measure different expansion rates (and therefore different deceleration parameters) from those following the background Hubble flow. We then explain how weak peculiar motions can lead to locally accelerated expansion within a globally decelerating universe. Section 3 provides the conditions for apparent universal acceleration and uses current observations to estimate the scale of the accelerated domain and the average value of the deceleration parameter there. In section 4 we consider anisotropic peculiar flows and obtain expressions for the main kinematic variables. These formulae are then used to discuss dipole-like patterns in the distribution of the deceleration parameter, which are caused by the peculiar motion itself. Finally, we close with a brief summary of our results and a general discussion.

\section{Peculiar kinematics in perturbed FRW
universes}\label{sPKPFRWUs}
The frame of the Microwave Background, namely the coordinate system where the CMB dipole vanishes, is believed to coincide with the frame of the smooth Hubble expansion. This is the preferred cosmological coordinate system, with respect to which large-scale peculiar velocities can be defined and measured.

\subsection{Expansion rates}\label{ssERs}
Suppose that $u_a$ is the reference 4-velocity of the CMB, relative to which the universe is an exact dust-dominated FRW model. In the perturbed spacetime, a typical observer in a galaxy like our Milky Way has
\begin{equation}
\tilde{u}_a= u_a+v_a\,,  \label{eq:tua}
\end{equation}
with $v_a$ representing the observer's drift velocity.\footnote{The $\tilde{u}_a$-field is also timelike irrespective of the magnitude of the peculiar velocity. Each frame defines its own time direction and 3-space. The tensors $h_{ab}=g_{ab}+u_au_b$ and  $\tilde{h}_{ab}=g_{ab}+ \tilde{u}_a\tilde{u}_b$, with $g_{ab}$ representing the spacetime metric, project orthogonal to $u_a$ and $\tilde{u}_a$ respectively. They also define the orthogonally projected covariant derivative operators, by means of ${\rm D}_a=h_a{}^b\nabla_b$ and $\tilde{\rm D}_a= \tilde{h}_a{}^b\nabla_b$ ($\nabla_a$ is the standard covariant derivative)~\cite{TCM}.} Note that $u_av^a=0$ always and $v^2=v_av^a\ll1$ in our case. The latter constraint guarantees that $\gamma=(1-v^2)^{-1/2}\simeq1$~\cite{KE}.

The mean kinematics of the peculiarly moving, the `tilded', observers are determined by the volume scalar ($\tilde{\Theta}= \nabla^a\tilde{u}_a$) of their worldline congruence~\cite{TCM}. Positive values for $\tilde{\Theta}$ imply that the average separation between these observers increases and indicate expansion. In the opposite case we have contraction. Similarly, $\Theta$ (with $\Theta=\nabla^au_a>0$) monitors the expansion of the universe. To first order in $v_a$, we have\footnote{Recall that we have assumed non-relativistic peculiar velocities, which implies that we can drop terms of order $v^2$ and higher from Eq.~(\ref{eq:Thetas}) and the rest of our formulae. We also use geometrised units with $c=1=8\pi G$.}
\begin{equation}
\tilde{\Theta}= \Theta+ \vartheta\,, \label{eq:Thetas}
\end{equation}
where $\vartheta=\tilde{\rm D}^av_a$. This scalar measures the mean separation between neighbouring peculiar-flow lines and here it will always satisfy the $\vartheta/\Theta\ll1$ condition. In regions where $\vartheta$ is positive, the peculiar motion adds to background expansion and the drifting observers expand faster than the actual universe (i.e.~$\tilde{\Theta}>\Theta$). Here, we will always assume positive $\vartheta$.\footnote{Systematically negative values of $\vartheta$ are usually associated with structures that have decoupled from the background expansion, `turned around' and started to collapse. Such gravitationally bound systems, however, have typical dimensions much smaller than those considered here.}

In multi-component systems each group of observers has its own time-direction. So, in our case, time can be measured relative to the CMB and along the tilded frame. Thus,  differentiating Eq.~(\ref{eq:Thetas}) with respect to time, we arrive at
\begin{equation}
\dot{\tilde{\Theta}}= \Theta^{\prime}+ \dot{\vartheta}\,,  \label{eq:dThetas}
\end{equation}
where overdots and primes indicate time derivatives along the $\tilde{u}_a$ and the $u_a$ fields respectively (i.e.~$\dot{\tilde{\Theta}}=\tilde{u}^a\nabla_a\tilde{\Theta}$, $\dot{\vartheta}=\tilde{u}^a\nabla_a\vartheta$ and $\Theta^{\prime}=u^a\nabla_a\Theta$). Also, in what follows, we will always assume that $\Theta^{\prime}<0$ and that $|\dot{\vartheta}/\Theta^{\prime}|\ll1$. The scalars $\dot{\tilde{\Theta}}$ and $\Theta^{\prime}$ measure changes in the expansion rates along each of the aforementioned two time-directions and are determined by the associated Raychaudhuri equations~\cite{TCM}. In a perturbed dust-dominated FRW model with small drift velocities, the Raychaudhuri formulae in the CMB and the tilded frames are
\begin{equation}
\Theta^{\prime}= -{1\over3}\,\Theta^2- {1\over2}\,\rho \hspace{15mm} {\rm and} \hspace{15mm} \dot{\tilde{\Theta}}= -{1\over3}\,\tilde{\Theta}^2- {1\over2}\,\tilde{\rho}+ \tilde{\rm D}^a\tilde{A}_a\,,  \label{eq:Rays1}
\end{equation}
respectively. Note that $\rho$ is the matter density relative to the $u_a$-field and $\tilde{\rho}$ represents its tilded counterpart, with $\tilde{\rho}=\rho$ to linear order in $v_a$~\cite{M}. Also, $\tilde{A}_a$ is the 4-acceleration measured by the drifting observer. This vector vanishes in the CMB frame (i.e.~$A_a=0$) but is nonzero in every other relatively moving coordinate system. In particular, $\tilde{A}_a= \dot{v}_a+(\Theta/3)v_a$, to linear order in $v_a$~\cite{M,ET}. The  4-acceleration term in Eq.~(\ref{eq:Rays1}b) is central to our analysis. Its presence means that expressions (\ref{eq:Rays1}a) and (\ref{eq:Rays1}b) are different, even when matter is pressureless dust and the peculiar velocities are small.\footnote{The fact that the scalars $\dot{\tilde{\Theta}}$ and $\Theta^{\prime}$ are generally different is also demonstrated in Eq.~(\ref{eq:dThetas}). Expressions (\ref{eq:Rays1}), however, provide additional information showing where and how this difference comes from.} In other words, observers drifting with respect to the CMB frame have expansion rates different from that of the actual universe simply because of their relative motion. For our purposes, this fact represents the most significant theoretical deviation from the conventional single-fluid studies.

\subsection{Deceleration parameters}\label{ssDPs}
If the expansion rate of a typical observer in a dust-dominated, almost-FRW universe differs from that of the Hubble flow, the deceleration parameters in the two frames should be different as well. One might then ask whether it is possible to experience accelerated expansion in one frame and decelerated in the other~\cite{T1}. Whether, in particular, the peculiarly moving observer could measure a negative deceleration parameter, while the universe is actually decelerating.

To investigate this possibility, we first need to write down the deceleration parameters as `measured' in the $u_a$ and $\tilde{u}_a$ frames. Expressed in terms of their corresponding volume scalars, these read
\begin{equation}
q= -\left(1+{3\Theta^{\prime}\over\Theta^2}\right) \hspace{15mm} {\rm and} \hspace{15mm} \tilde{q}= -\left(1+{3\dot{\tilde{\Theta}}\over\tilde{\Theta}^2}\right)\,, \label{eq:qs1}
\end{equation}
respectively. Substituting the above into Eqs.~(\ref{eq:Rays1}), the latter recast into
\begin{equation}
(1+q)\Theta^2= \Theta^2+ {3\over2}\,\rho \hspace{15mm} {\rm and} \hspace{15mm} (1+\tilde{q})\tilde{\Theta}^2= \tilde{\Theta}^2+ {3\over2}\,\tilde{\rho}- 3\tilde{\rm D}^a\tilde{A}_a\,.  \label{eq:Rays2}
\end{equation}
Although these expressions already reveal that $q$ and $\tilde{q}$ are generally different, it helps to relate the two deceleration parameters directly. Following~\cite{T1}, we recall that $\tilde{\rho}=\rho$ and $\tilde{A}_a=\dot{v}_a+ (\Theta/3)v_a$ to first order in $v_a$. Then, we can combine Eqs.~(\ref{eq:Rays2}) to a single relation by eliminating the matter density. In particular, using the definition $\vartheta=\tilde{\rm D}^av_a$, together with expression (\ref{eq:Thetas}) and the (linear in $v_a$) relation $\dot{\vartheta}=\tilde{\rm D}^a\dot{v}_a -\Theta\vartheta/3$ (see~\cite{ET}), we arrive at
\begin{equation}
1+ \tilde{q}= (1+q)
\left(1+{\vartheta\over\Theta}\right)^{-2}-
{3\dot{\vartheta}\over\Theta^2} \left(1+{\vartheta\over\Theta}\right)^{-2}\,,  \label{eq:tq2}
\end{equation}
where $\Theta$, $\vartheta>0$ and $\vartheta/\Theta\ll1$ always. Note that the above result can be also obtained from Eq.~(\ref{eq:dThetas}) by means of (\ref{eq:Thetas}) and (\ref{eq:qs1})~\cite{T1}.

Expression (\ref{eq:tq2}) provides the deceleration parameter ($\tilde{q}$), as measured by an observer drifting with respect to the smooth Hubble flow of a dust-dominated FRW model, relative to that of the universe itself ($q$). Following (\ref{eq:tq2}), it is clear that $q$ and $\tilde{q}$ differ in general. Moreover, as long as the right-hand side of Eq.~(\ref{eq:tq2}) remains below unity, positive values for $q$ do not a priori guarantee the same for $\tilde{q}$. It is therefore theoretically possible for the tilded observer to experience accelerated expansion in a decelerating universe. This, however, can only happen locally. Indeed, assuming that the universe approaches an exact FRW model on large enough scales, peculiar velocities and their effects should die away as we move on to progressively larger lengths. In other words, there should always be a threshold beyond which $\vartheta=0=\dot{\vartheta}$ and $\tilde{q}=q$. The question is how far out such a threshold could be (see \S~\ref{ssLAE} later).

\subsection{Weakly and strongly accelerated
expansion}\label{ssWSAE}
Before closing this section, it is worth noting that there are two types of accelerated expansion, depending on the value of the deceleration parameter. In particular, the condition $-1<\tilde{q}<0$ is equivalent to $-\tilde{\Theta}^2/3< \dot{\tilde{\Theta}}<0$, which means that $\tilde{q}$ and $\dot{\tilde{\Theta}}$ can be simultaneously negative. Naturally, analogous relations also hold between $q$, $\Theta^2$ and $\Theta^{\prime}$ (see (\ref{eq:qs1})), as well as between the corresponding variables of the peculiar motion. On these grounds, it helps to distinguish between accelerated expansion with simply $-1<\tilde{q}<0$ and that with $\dot{\tilde{\Theta}}>0$ (namely with $\tilde{q}<-1$). We will therefore associate condition $-1<\tilde{q}<0$ with {\em weakly accelerated expansion} and treat $\dot{\tilde{\Theta}}>0$ (equivalently $\tilde{q}<-1$) as the necessary requirement for {\em strongly accelerated expansion}.

The distinction between the aforementioned two types of acceleration is important because, although two decelerating expansions cannot lead to a strongly accelerating one, they can combine to give weak acceleration. This means that, although both the Hubble flow and the peculiar motion can be decelerating, the overall expansion of the drifting observer can be weakly accelerating. To be precise, suppose that $\Theta^{\prime}<-\Theta^2/3$ and $\dot{\vartheta}<-\vartheta^2/3$, to ensure that the background expansion and the peculiar flow are both decelerating. Then, starting from Eq.~(\ref{eq:dThetas}), we find that $\dot{\tilde{\Theta}}<-(\Theta^2+\vartheta^2)/3<0$, which excludes the possibility of a strongly accelerated overall motion. Nevertheless, on using (\ref{eq:Thetas}), the same expression recasts into
\begin{equation}
\dot{\tilde{\Theta}}<-{1\over3}\,\tilde{\Theta}^2+ {2\over3}\,\Theta\vartheta\,,  \label{eq:waccel1}
\end{equation}
with $\Theta$, $\vartheta>0$ in our case. Clearly, this result allows for the possibility that
\begin{equation}
-{1\over3}\,\tilde{\Theta}^2<\dot{\tilde{\Theta}}<0\,,  \label{eq:waccel2}
\end{equation}
implying $-1<\tilde{q}<0$ for the tilded observer. In other words, peculiar motions can lead to weakly accelerated expansion within the limits of the linear (the almost-FRW) approximation.~\footnote{Perhaps the most direct way of demonstrating that two decelerating expansions can lead to an accelerated one, is by writing Eq.~(\ref{eq:Thetas}) as $\dot{\tilde{a}}/\tilde{a}= (\dot{a}/a)+(\dot{\alpha}/\alpha)$, where $\tilde{a}$, $a$ and $\alpha$ are the three scale factors (with $\dot{a},\,\dot{\alpha}>0$). Then, $\ddot{\tilde{a}}/\tilde{a}= (\ddot{a}/a)+(\ddot{\alpha}/\alpha)+ 2(\dot{a}/a)(\dot{\alpha}/\alpha)$, meaning that negative values for $\ddot{a}$ and $\ddot{\alpha}$ do not a priori guarantee the same for $\ddot{\tilde{a}}$. Note that for simplicity we have used overdots for both time derivatives.} Noting that the supernovae results point towards weak acceleration, since they put the deceleration parameter close to $-0.5$~\cite{TR}, we will focus on the $-1<\tilde{q}<0$ case for the rest of this report.

\section{Apparent acceleration due to peculiar
motions}\label{aAAPMs}
The main theoretical principle following from our discussion so far, is that measuring a negative deceleration parameter in a frame drifting relative to the CMB (like that of our Milky Way for example) may simply be a local effect and does not necessarily imply a globally accelerating universe. Our next step is to examine the conditions for this to happen and to what extent current observations could support such a scenario.

\subsection{Conditions for local acceleration}\label{ssCLA}
Observers will experience local acceleration, as a direct result of their own peculiar motion relative to the smooth Hubble flow, in a region where the right-hand side of Eq.~(\ref{eq:tq2}) drops below unity. Recalling that in the absence of matter pressure $q=\Omega/2$ and $3\Theta^{\prime}= -\Theta^2[1+(\Omega/2)]$, the latter expression recasts into
\begin{equation}
1+ \tilde{q}= \left(1+{1\over2}\,\Omega\right)
\left(1+{\vartheta\over\Theta}\right)^{-2} \left(1+{\dot{\vartheta}\over\Theta^{\prime}}\right)\,,  \label{eq:tq3}
\end{equation}
where $\Omega$, $\Theta$, $\vartheta>0$ and $\Theta^{\prime}<0$ always. Then, noting that $\Omega$ may be seen as the effective density parameter of the region in question, rather than that of the universe itself, the condition for locally accelerated expansion reads
\begin{equation}
\left(1+{1\over2}\,\Omega\right)
\left(1+{\vartheta\over\Theta}\right)^{-2} \left(1+{\dot{\vartheta}\over\Theta^{\prime}}\right)< 1\,.  \label{eq:cla}
\end{equation}

So far, we have assumed drift velocities with $v^2\ll1$ in a dust-dominated, almost-FRW universe. It should be noted here that the proximity to the FRW model has not followed from geometrical considerations imposed on the perturbed metric and its derivatives. Instead, in line with the covariant nature of our approach, the near-FRW claim follows from the conditions imposed on the kinematic scalars and by demanding that the homogeneity scale of the universe lies inside the horizon. In particular, to ensure that the peculiar kinematics are always subdominant to the Hubble flow we have demanded that $\vartheta/\Theta$, $|\dot{\vartheta}/\Theta^{\prime}|\ll1$ at all times (see \S~\ref{ssERs} earlier). Although $\vartheta$ is always assumed positive, however, its time derivative can (in principle at least) take either sign. Qualitatively speaking, positive values for $\dot{\vartheta}$ will assist the (local) acceleration, while negative ones will do the opposite. We may therefore distinguish between the following three characteristic cases:

\begin{itemize}

\item Suppose that $\dot{\vartheta}<0$ and $0<\dot{\vartheta}/\Theta^{\prime}\ll\vartheta/\Theta$ (recalling that $\Theta^{\prime}<0$ always). In this occasion, which one could argue that corresponds to peculiar motions with very slowly varying relative expansion rates, condition (\ref{eq:cla}) reduces to
    \begin{equation}
    \left(1+{1\over2}\,\Omega\right)
    \left(1+{\vartheta\over\Theta}\right)^{-2}< 1\,.  \label{eq:cla1}
    \end{equation}
    At the same time, the associated deceleration parameter (see Eq.~(\ref{eq:tq3})) is given by
    \begin{equation}
    \tilde{q}= \tilde{q}_{\star}\simeq \left(1+{1\over2}\,\Omega\right)
    \left(1+{\vartheta\over\Theta}\right)^{-2}- 1\,.  \label{eq:tq3i}
    \end{equation}

\item Alternatively, we may assume that $\dot{\vartheta}<0$ with $0<\dot{\vartheta}/\Theta^{\prime}\simeq\vartheta/\Theta\ll1$. Then, the condition for locally accelerated expansion strengthens to
    \begin{equation}
    \left(1+{1\over2}\,\Omega\right)
    \left(1+{\vartheta\over\Theta}\right)^{-1}< 1\,,  \label{eq:cla2}
    \end{equation}
    with
    \begin{equation}
    \tilde{q}= \tilde{q}_{\dagger}\simeq \left(1+{1\over2}\,\Omega\right)
    \left(1+{\vartheta\over\Theta}\right)^{-1}- 1\,.  \label{eq:tq3ii}
    \end{equation}

\item Finally, when $\dot{\vartheta}>0$, with $\dot{\vartheta}/\Theta^{\prime}<0$ and $|\dot{\vartheta}/\Theta^{\prime}|\simeq\vartheta/\Theta\ll1$, condition (\ref{eq:cla}) relaxes to
    \begin{equation}
    \left(1+{1\over2}\,\Omega\right)
    \left(1+{\vartheta\over\Theta}\right)^{-2} \left(1-{\vartheta\over\Theta}\right)< 1  \label{eq:cla3}
    \end{equation}
    and the corresponding value of the deceleration parameter is determined by
    \begin{equation}
    \tilde{q}= \tilde{q}_{\ddagger}\simeq \left(1+{1\over2}\,\Omega\right)
    \left(1+{\vartheta\over\Theta}\right)^{-2} \left(1-{\vartheta\over\Theta}\right)- 1\,.  \label{eq:tq3iii}
    \end{equation}

\end{itemize}
The last of the aforementioned conditions is the the most favourable of the three, as far as local acceleration is concerned, while the second is the least favourable one. The first condition, on the other hand, corresponds to an intermediate situation and for this reason the coming sections will mainly focus on this case.

Clearly, in regions where the right-hand side of (\ref{eq:cla1}), or of (\ref{eq:cla2}) and (\ref{eq:cla3}), drops below unity, the drifting observers will experience locally accelerated expansion in a globally decelerating universe. Whether this happens or not, as well as the scale of the affected area, depends on the speed of the drift flow and on the density of the region in question. In particular the faster the peculiar motion and the lower the density, the larger the accelerated patch and the lower (more negative) the deceleration parameter. So, the next question is whether observations allow for peculiar velocities large enough to trigger such apparent acceleration.

\subsection{Locally accelerated expansion}\label{ssLAE}
\begin{figure}[tbp]
\centering \vspace{7cm} \includegraphics{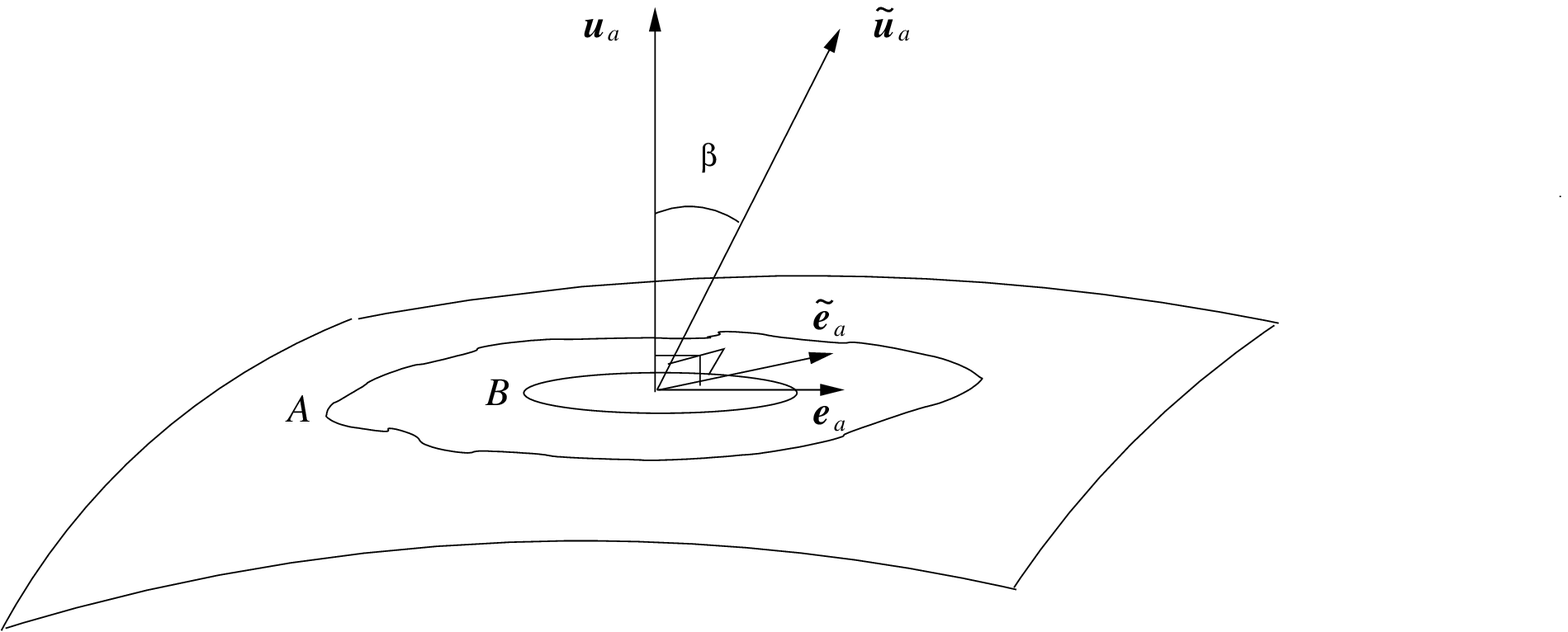} \caption{Typical observers in section $\mathcal{A}$ move along the $\tilde{u}_a$-field, which drifts relative to the CMB frame ($u_a$) with peculiar velocity $v_a$. The patch $\mathcal{A}$ has positive $\vartheta=\tilde{\rm D}^av_a$ and so expands faster than the background universe (see Eq.~(\ref{eq:Thetas})). Inside region $\mathcal{B}$ the right-hand side of expression (\ref{eq:tq3}) drops below unity and there the `tilded' observer experiences accelerated expansion~\cite{T1}.}  \label{fig:pvel}
\end{figure}

Consider an expanding spatial region $\mathcal{A}$ -- see Fig.~\ref{fig:pvel}, which largely complies with the FRW symmetries, but is still endowed with a weak, bulk peculiar velocity field that `adds' to the background expansion (i.e.~$\vartheta>0$, with $\vartheta/\Theta\ll1$). Typical observers in $\mathcal{A}$ have peculiar velocities close to the bulk flow of the patch (both in magnitude and direction) and their expansion rate is determined by Eq.~(\ref{eq:tq2}). According to the latter, the deceleration parameter measured by the tilded observer is generally different from that of the background universe. In particular, around any given observer in $\mathcal{A}$ there can be a region $\mathcal{B}$ where the right-hand side of (\ref{eq:tq2}) -- see also (\ref{eq:tq3}) -- drops below unity and where the deceleration parameter takes negative values. As we will see below, the size of the accelerated domain $\mathcal{B}$ and the value of the deceleration parameter there, depend on the specifics of the peculiar expansion and on the density distribution of the matter within section $\mathcal{A}$.

Peculiar velocities are difficult to assess directly. As a result, the speed of bulk drift flows is usually estimated by means of statistical methods (e.g.~see~\cite{SW}). The current concordance cosmological model, the $\Lambda$CDM, allows for drift flows no faster than $\sim100$~km/sec on scales close to 100~Mpc that drop further as we move out to larger lengths. Recent independent surveys, however, have reported significantly faster bulk velocities on substantially greater scales~\cite{KA-BKE}-\cite{LTMC}, putting the standard picture in doubt~\cite{P}. The surveys of~\cite{KA-BKE}, in particular, indicate peculiar velocities as fast as $\sim1000$~km/sec extending out to $\sim800$~Mpc, and perhaps all the way to the Hubble radius.\footnote{The surveys extend to 100$h^{-1}$Mpc~\cite{WFH,LTMC}, 300$h^{-1}$~Mpc and 500$h^{-1}$~Mpc~\cite{KA-BKE}, showing bulk velocities as large as 500~km/sec and 1000~km/sec respectively. On smaller lengths (between 30 and 60~Mpc) the work of~\cite{LS} suggests a (positive) variance in the local Hubble rate of up to 10\%. With the possible exception of the last survey, there is currently no way of knowing whether these bulk flows are of the desired type (i.e.~with $\vartheta>0$). Nevertheless, in the absence of better data, we will use the above measurements to infer order-of-magnitude estimates for $\vartheta$.} These values translate into $\vartheta/\Theta\simeq1/7$ on scales close to 100~Mpc  (assuming that $H\simeq70$~km/secMpc today).\footnote{We use the approximate relation $\vartheta=\tilde{\rm D}_av^a\simeq \partial v^a/\partial r^a\sim3v/r$, where $v$ is the magnitude of the mean bulk velocity and $r$ the average size of the associated region. It is then straightforward to show that $\vartheta/\Theta\sim v/Hr$, with $\Theta=3H$ and $H$ representing the background Hubble parameter.} There, for an (effective) density parameter $\Omega\simeq1/25$, expressions (\ref{eq:tq3i})-(\ref{eq:tq3iii}) lead to
\begin{equation}
\tilde{q}_{\star}\simeq-0.22\,, \hspace{20mm} \tilde{q}_{\dagger}\simeq-0.11 \hspace{15mm} {\rm and} \hspace{15mm} \tilde{q}_{\ddagger}\simeq-0.33\,,  \label{eq:tqs1}
\end{equation}
respectively. In every case the deceleration parameter, as measured in the tilded frame, is negative. This means that the drifting observers experience locally accelerated expansion within a globally decelerating universe, as a result of their peculiar motion alone.

As we move on to larger regions the relative magnitude of the drift velocity drops, the acceleration effect becomes progressively less prominent and eventually fades away. Nevertheless, depending on the value of the peculiar velocity and on the density of the region in question, one could reach as far out as several hundred (or even a few thousand) Mpc, before the smooth Hubble flow is restored. For instance, on lengths close to 500~Mpc, the surveys of~\cite{KA-BKE} suggest that $\vartheta/\Theta\simeq1/35$. There, maintaining the same value for the density parameter (i.e.~$\Omega\simeq1/25$), Eqs.~(\ref{eq:tq3i})-(\ref{eq:tq3iii}) give
\begin{equation}
\tilde{q}_{\star}\simeq-0.036\,, \hspace{20mm} \tilde{q}_{\dagger}\simeq-0.008 \hspace{15mm} {\rm and} \hspace{15mm} \tilde{q}_{\ddagger}\simeq-0.063\,,  \label{eq:tqs2}
\end{equation}
respectively.\footnote{The more modest drift velocities reported by~\cite{WFH,LTMC} also lead to local acceleration, though then the accelerated region is generally smaller and the deceleration parameter less negative (see~\cite{T1} for numerical examples).} Further out, close to 1000~Mpc for example, the relative speed of the peculiar motion drops down to $\vartheta/\Theta\simeq1/70$. There, assuming that $\Omega\simeq1/100$ and using expressions (\ref{eq:tq3i})-(\ref{eq:tq3iii}), we find that
\begin{equation}
\tilde{q}_{\star}\simeq-0.023\,, \hspace{20mm} \tilde{q}_{\dagger}\simeq-0.009 \hspace{15mm} {\rm and} \hspace{15mm} \tilde{q}_{\ddagger}\simeq-0.037\,.  \label{eq:tqs3}
\end{equation}
So, in this scenario, the accelerated expansion is only an apparent, local effect and beyond a certain scale-threshold the background Hubble flow is restored. To the unsuspecting observer, however, it may look as though the whole universe has moved into a phase of recent acceleration.

\section{Anisotropy in the $\tilde{q}$-distribution}\label{sAQD}
So far, we have focused on the average kinematics of peculiar motions, without incorporating anisotropies into our analysis. These can be triggered by a number of agents and in principle they could lead to observable anisotropy patterns in the spatial distribution of the deceleration parameter, as measured by the drifting observer.

\subsection{Anisotropic expansion rates}\label{ssAERs}
To discuss deviations from isotropy in the spatial distribution of the deceleration parameter, which are themselves induced by anisotropies in the observer's peculiar motion, we first need to consider the kinematics of anisotropically expanding media. In our case all the scales of interest are well within the (possible) curvature radius of the universe. Given that, we may ignore any general relativistic effects. Then, in the absence of vorticity, the generalised Hubble's law reads
\begin{equation}
\Theta_{\alpha\beta}= {1\over3}\,\Theta\delta_{\alpha\beta}+ \sigma_{\alpha\beta}\,,  \label{eq:Thab}
\end{equation}
where $\Theta_{\alpha\beta}=\partial_{(\beta}u_{\alpha)}$ and $\sigma_{\alpha\beta}=\partial_{(\beta}u_{\alpha)}- (\Theta/3)\delta_{\alpha\beta}$ -- with $\sigma_{\alpha}{}^{\alpha}=0$ -- represent the expansion and the shear tensors respectively.\footnote{Greek indices run from 1 to 3, while the Latin ones (used in the previous sections) take their values between 0 and 3. Also, round brackets denote symmetrisation and square ones indicate antisymmetrisation.} In addition, $\Theta=\partial^{\alpha}u_{\alpha}$ and $\delta_{\alpha\beta}$ is the familiar Kronecker delta. Applied along the three principle shear eigen-directions, the above relation gives
\begin{equation}
\Theta_{\alpha}= {1\over3}\,\Theta+ \sigma_{\alpha}\,,  \label{eq:Thalp}
\end{equation}
with $\sum_{\alpha=1}^3\sigma_{\alpha}=0$ and $\sum_{\alpha=1}^3\Theta_{\alpha}=\Theta$. As expected, positive shear values along a given spatial direction enhance the expansion, while negative ones slow it down. One can also use the diagonal components of the expansion tensor to define the effective scale factors in the aforementioned three eigen-directions by means of $\dot{a}_{\alpha}/a_{\alpha}=\Theta_{\alpha}$. Then, expression (\ref{eq:Thalp}) translates into
\begin{equation}
{\dot{a}\over a}= {1\over3}\left({\dot{a}_1\over a_1} +{\dot{a}_2\over a_2}+{\dot{a}_3\over a_3}\right)\,,  \label{eq:alphas}
\end{equation}
which relates the average with the direction-dependent scale factors. Furthermore, recalling that $q=-\ddot{a}a/\dot{a}^2$ is the average deceleration parameter, we define its direction dependent counterparts as
\begin{equation}
q_{\alpha}= -{\ddot{a}_{\alpha}a_{\alpha}\over\dot{a}_{\alpha}^2}= -\left(1+{\dot{\Theta}_{\alpha}\over\Theta_{\alpha}^2}\right)\,.  \label{eq:anqs}
\end{equation}
With this definition in hand, it is simple to obtain the relation between the average and the direction-dependent deceleration parameters. In particular, taking the time-derivative of Eq.~(\ref{eq:alphas}), using definition (\ref{eq:anqs}) and then employing some fairly straightforward algebra, we arrive at
\begin{eqnarray}
1+ q&=& 3\left[(1+q_1)\left({\Theta_1\over\Theta}\right)^2 +(1+q_2)\left({\Theta_2\over\Theta}\right)^2 +(1+q_3)\left({\Theta_3\over\Theta}\right)^2\right] \nonumber\\ &=&{1\over3}\left[(1+q_1)\left(1+{3\sigma_1\over\Theta}\right)^2 +(1+q_2)\left(1+{3\sigma_2\over\Theta}\right)^2 +(1+q_3)\left(1+{3\sigma_3\over\Theta}\right)^2\right]\,.  \label{eq:qvsqs}
\end{eqnarray}
Note that the second equality has been obtained from the first after using relation (\ref{eq:Thalp}). Also, in the case of zero shear anisotropy, namely when $\sigma_1=\sigma_2=\sigma_3=0$ and $q_1=q_2=q_3=q$, the above relation reduces to a trivial identity.

\subsection{Shear-like anisotropy}\label{ssS-LA}
Anisotropies in the observed spatial distribution of the deceleration parameter can appear for a number of reasons and in a variety of ways. For instance, the symmetry of domain $\mathcal{A}$ (see Fig.~\ref{fig:pvel}) and the observer's position in it can induce anisotropy in the observed $\tilde{q}$-distribution. This type of anisotropy, however, is less of an issue when the patch $\mathcal{A}$ is considerably larger than $\mathcal{B}$. In that case only observers `living' near the edge of region $\mathcal{A}$ will be affected. Further deviation from isotropy can occur because the peculiar flow itself is not shear-free, or because the drifting observers have their own individual peculiar velocities, which are generally different from each other and also from the mean bulk motion of the patch.

When shear anisotropy is included, the expansion tensor (see expression (\ref{eq:Thab})) associated with the overall motion of the drifting (the tilded) observer reads
\begin{equation}
\tilde{\Theta}_{\alpha\beta}= {1\over3}\,\Theta\delta_{\alpha\beta}+ \vartheta_{\alpha\beta}= {1\over3}\left(\Theta +\vartheta\right)\delta_{\alpha\beta}+ \sigma_{\alpha\beta}\,,  \label{eq:tThab}
\end{equation}
where $\tilde{\Theta}_{\alpha\beta}= \partial_{(\beta}\tilde{u}_{\alpha)}$, $\Theta= \partial^{\alpha}u_{\alpha}$ and $\vartheta_{\alpha\beta}= \partial_{(\beta}v_{\alpha)}=(\vartheta/3)\delta_{\alpha\beta}+ \sigma_{\alpha\beta}$. The latter is the expansion tensor of the drift motion, with $\vartheta=\partial^{\alpha}v_{\alpha}$ representing the associated volume scalar and $\sigma_{\alpha\beta}= \partial_{(\beta}v_{\beta)}- (\vartheta/3)\delta_{\alpha\beta}$ the peculiar shear (with $\sigma^{\alpha}{}_{\alpha}=0$). Also, as in \S~\ref{ssAERs} before, we have assumed a non-rotating peculiar velocity field and set $\partial_{[b}v_{a]}$ to zero. Clearly, in the absence of anisotropy, $\sigma_{\alpha\beta}=0$ and the above relation reduces to Eq.~(\ref{eq:Thetas}). Moreover, written along the three principle shear eigen-directions, expression (\ref{eq:tThab}) recasts into
\begin{equation}
\tilde{\Theta}_{\alpha}= {1\over3}\,\Theta+ \vartheta_{\alpha}= {1\over3}\left(\Theta+\vartheta\right)+ \sigma_{\alpha}= {1\over3}\,\tilde{\Theta}+ \sigma_{\alpha}\,,  \label{eq:tThi}
\end{equation}
with $\sum_{\alpha=1}^3\sigma_{\alpha}=0$ and $\sum_{\alpha=1}^3\vartheta_{\alpha}=3\vartheta$. We see again that positive/negative shear values along a given direction will enhance/reduce the overall expansion there. Taking the time derivative of the above, using definitions (\ref{eq:qs1}a) and (\ref{eq:anqs}), together with the background relations $q=\Omega/2$ and $\Theta^{\prime}=(\Theta^2/3)(1+\Omega/2)$, we arrive at
\begin{equation}
1+\tilde{q}_{\alpha}= \left(1+{1\over2}\,\Omega\right) \left(1+{\vartheta\over\Theta} +3\,{\sigma_{\alpha}\over\Theta}\right)^{-2} \left[1+{\dot{\vartheta}\over\Theta^{\prime}} \left(1+3\,{\dot{\sigma}_{\alpha}\over\dot{\vartheta}}\right)\right]\,, \label{eq:antqs}
\end{equation}
where we remind the reader that $\Omega$ may be seen as the effective density parameter of section $\mathcal{A}$. This formula provides the deceleration parameter measured by the tilded observer, in the case of a generally anisotropic peculiar motion, along the principal shear eigen-directions. When combined with Eq.~(\ref{eq:qvsqs}), the above leads to the average expression (\ref{eq:tq2}), thus verifying the mathematical consistency of our analysis.

As mentioned at the beginning of this section, anisotropies in the $\tilde{q}$-distribution can emerge from a variety of reasons. In what follows, we will use relation (\ref{eq:antqs}) to estimate apparent dipole-like patterns in the spatial distribution of the deceleration parameter that result from the observer's peculiar motion alone.

\subsection{Dipole-like anisotropy}\label{ssD-LA}
Of all the possible forms of anisotropy, the one most typically associated with peculiar motions is probably of the dipole type, caused by the fact that the drift flow introduces a preferred direction to the observer's three dimensional space. This in turn gives rise to a characteristic (apparent) dipolar axis, relative to a given `rest frame'. The CMB dipole, in particular, is believed to be such a Doppler-like effect, reflecting the peculiar motion of our Local Group relative to the rest-frame of the smooth Hubble flow. Similarly, differences between the peculiar velocity of an observer lying somewhere within section $\mathcal{A}$ and the mean flow of that patch, should lead to an analogous dipole-like pattern in the spatial distribution of the deceleration parameter in that region.

Consider a typical observer within $\mathcal{A}$, drifting with peculiar velocity close to the bulk flow of the patch.\footnote{Throughout this report we only consider the kinematics of typical observers. By definition, these have peculiar velocities close, both in magnitude and in direction, to the mean bulk flow of the drifting domain.} Without loss of generality, we may also assume that the observer moves along the first of the three shear eigen-directions (see \S~\ref{ssS-LA} before). To estimate the apparent dipolar anisotropy that the peculiar motion induces into the $\tilde{q}$-distribution, we will assign a positive value to the apparent shear ($\sigma_1^{(+)}$) along the observer's motion. Then, in the opposite direction, the corresponding apparent shear will be $\sigma_1^{(-)}=-\sigma_1^{(+)}$. Given that our observers are typical and that the aforementioned peculiar shear simply reflects their motion with respect to the mean bulk flow of the region (i.e.~to the rest-frame of patch $\mathcal{A}$), the associated anisotropy should be small. In other words, $\sigma_1^{(\pm)}/\vartheta\ll1$. Here, mainly for illustration purposes, we assume that $\sigma_1^{(+)}=\vartheta/15= -\sigma_1^{(-)}$. Then, expression (\ref{eq:antqs}) gives
\begin{equation}
\tilde{q}_1^{(+)}= \left(1+{1\over2}\,\Omega\right) \left(1+{6\over5}\,{\vartheta\over\Theta}\right)^{-2} \left(1+{6\over5}\,{\dot{\vartheta}\over\Theta^{\,\prime}}\right)- 1\,,  \label{eq:santq1+}
\end{equation}
towards the direction of the motion and
\begin{equation}
\tilde{q}_1^{(-)}= \left(1+{1\over2}\,\Omega\right) \left(1+{4\over5}\,{\vartheta\over\Theta}\right)^{-2} \left(1+{4\over5}\,{\dot{\vartheta}\over\Theta^{\,\prime}}\right)- 1\,,  \label{eq:santq1-}
\end{equation}
in the opposite way. Applying the above to a region of approximately 100~Mpc, where $\vartheta/\Theta\simeq1/7$ and $\Omega\simeq1/25$, while setting $\dot{\vartheta}/\Theta^{\prime}\ll\vartheta/\Theta$, we obtain
\begin{equation}
\tilde{q}_1^{(+)}\simeq -0.26 \hspace{15mm} {\rm and} \hspace{15mm} \tilde{q}_1^{(-)}\simeq -0.18\,.  \label{eq:tqdipole1}
\end{equation}
These results should be compared to the average value of $\tilde{q}_*\simeq-0.22$ obtained earlier for the same region and under the same assumptions (see Eq.~(\ref{eq:tqs1}a) in \S~\ref{ssLAE}). According to (\ref{eq:tqdipole1}), there is a small, but potentially measurable, dipole anisotropy in the spatial distribution of $\tilde{q}$ along the direction of the peculiar flow. In particular, the universe seems to accelerate faster in one direction (that of the peculiar motion -- where the apparent shear is positive). In the opposite direction, on the other hand, the acceleration seems to slow down by an equal amount (relative to the average).\footnote{What happens is that the whole of region $\mathcal{A}$ drifts with respect to the Hubble expansion, while every observer inside $\mathcal{A}$ also moves relative to the mean bulk flow of that patch. The former motion is essentially the one causing the CMB dipole. The latter leads to a dipolar anisotropy in the $\tilde{q}$-distribution within $\mathcal{A}$. For typical observers in $\mathcal{A}$, with drift velocities close to the average one, the $\tilde{q}$-dipole should be less pronounced than its CMB counterpart.} This pattern is maintained as we move further out to larger scales, where the effects of the peculiar motions weaken. On lengths close to 500~Mpc, for example, expressions (\ref{eq:santq1+}) and (\ref{eq:santq1-}) give
\begin{equation}
\tilde{q}_1^{(+)}\simeq -0.046 \hspace{15mm} {\rm and} \hspace{15mm} \tilde{q}_1^{(-)}\simeq -0.026\,,  \label{eq:tqdipole2}
\end{equation}
instead of the average $\tilde{q}_*=-0.036$ (see Eq.~(\ref{eq:tqs2}a) in \S~\ref{ssLAE}). Again, a small but in principle observable dipole anisotropy appears in the $\tilde{q}$-distribution.\footnote{One can easily extend results (\ref{eq:tqdipole1}) and (\ref{eq:tqdipole2}) to incorporate peculiar motions with $|\dot{\vartheta}/\Theta^{\prime}|\simeq\vartheta/\Theta$ (see \S~\ref{ssLAE}).}

One might also wonder about the orientation of the dipolar axis. This is a considerably more demanding theoretical task. Nevertheless, it sounds plausible to argue that the $\tilde{q}$-dipole should not lie far from its CMB counterpart. This is based on the fact that both dipoles are caused by the observer's peculiar flow. On the other hand, we should not expect the associated dipole axes to coincide either, since their corresponding reference frames are different. In the CMB case the rest-frame is that of the smooth Hubble expansion, while here it coincides with the average bulk flow of section $\mathcal{A}$. Recently, a number of reports seem to indicate that a small dipolar anisotropy, which is more-or-less aligned with the CMB dipole and is occasionally referred to as the ``cosmological axis'', might actually exist in the supernovae data~\cite{KL}.

\section{Discussion}\label{sD}
A little more than a decade ago, observations of high-redshift supernovae indicated that our universe was expanding at an accelerating pace~\cite{Retal}. This conclusion was reached after applying the supernovae luminosity distances to the distance-redshift relation of an exact FRW model. The results have repeatedly returned negative values for $q$, indicating an accelerated expansion for our universe. In particular, the deceleration parameter was estimated close to -0.5. The same measurements also suggested that the accelerated phase was a relatively recent event, putting the transition from deceleration to acceleration around $z=0.5$~\cite{TR}. As one looks at smaller redshifts, however, the picture seems to become less clear and there have been suggestions that the cosmic acceleration may have just peaked and started to slow down~\cite{ST}. Explaining the supernovae results has dominated almost every aspect of contemporary cosmology. Dark energy, an unknown and elusive form of matter with negative gravitational energy, has so far been the most popular answer. There has been scepticism, however, since the dark-energy scenario lacks a satisfactory explanation in terms of fundamental physics and also faces an awkward coincidence problem (see~\cite{S} for a discussion). Possible alternatives to dark energy, include modifying General Relativity, introducing extra dimensions, or abandoning the Friedmann models altogether.

The reason behind such drastic measures was that negative values for the deceleration parameter seemed impossible in conventional FRW cosmologies, as well as in perturbed FRW models. This is not necessarily the case however. Peculiar motions can locally induce weakly accelerated expansion (with $-1<\tilde{q}<0$ -- see \S~\ref{ssWSAE}), even when the drift velocities are small and matter is simple pressure-free dust, namely within the limits of the linear approximation. Moreover, although the effect is local, the affected scales can be large enough (of the order of several hundred Mpc -- perhaps even larger) to give the false impression that the whole universe has recently moved into a phase of global acceleration.

Technically speaking, the physics discussed here stems from the fact that the Raychaudhuri equations in the two coordinate systems (the CMB and that of a drifting observer) are not the same. The difference is in the 4-acceleration term, which vanishes in the CMB frame but takes nonzero values in any other relatively moving reference system. In practice, this means that the expansion rate of the drifting observer and that of the background Hubble flow are generally different. It is then theoretically possible to experience local acceleration in a globally decelerating universe. Alternatively, one could say that the observer's total motion is the `sum' of the smooth Hubble expansion and of their own peculiar flow. Then, it is fairly easy to show that the two constituent motions (although both decelerating) can in principle combine to give weakly accelerated expansion (with $-1<\tilde{q}<0$ -- see \S~\ref{ssWSAE}). The supernovae data seem to point towards weak acceleration.

There are, of course, certain requirements that need to be fulfilled, if peculiar motions are to cause local acceleration. The main one is that the drift flow should add to the Hubble expansion (i.e.~$\vartheta=\tilde{\rm D}^av_a>0$). Put another way, the patch $\mathcal{A}$ in Fig.~\ref{fig:pvel} should be expanding faster (even slightly faster) than the universe itself. In that case, the two decelerating expansions (the background and the peculiar) can locally combine to give a weakly accelerating one. One might then ask what is the likelihood for the Milky Way to rest within a region like $\mathcal{A}$. Given that the scales of interest are of cosmological relevance, the chances could be fairly high (perhaps higher than 50\%). The reason is that systematically negative values of $\vartheta$ are primarily associated with structures that have already decoupled from the background expansion, `turned around' and started to collapse. Such gravitationally bound systems, however, are believed to be considerably smaller in size.

The scale of the accelerated patch and the value of the deceleration parameter there, are decided by the speed of the drift flow and by the density of the matter in the section. In general, the larger the peculiar velocity and the lower the matter density, the larger the accelerated region and the lower (the more negative) the deceleration parameter. In our examples, we have adopted the peculiar velocities reported in the surveys of~\cite{KA-BKE}. These suggest drift flows of approximately 1000~km/sec extending as far out as 800~Mpc (perhaps even further out). Based on these, we found that the accelerated patch could reach lengths of the order of 1000~Mpc. The deceleration parameter was found to drop with scale, with its value ranging between $-0.11$ and $-0.33$ close to 100~Mpc and tending towards zero as one moves on to progressively larger lengths. As far as the matter density is concerned, we have assumed values in the $(0.01,0.1)$ range for the effective density parameter of the accelerated patch. The use of underdense regions, which expand faster than the background universe, renders certain analogies between this scenario and the large-void models~\cite{C}. Here, however, we do not need to reside near the centre of the void. Also, given that peculiar velocities are the result of structure formation, there are parallels with the `backreaction' scenarios as well (see~\cite{R} for a recent account).

Drift flows are typically associated with dipolar anisotropies, triggered by the preferred spatial direction that the peculiar velocities introduce. The CMB dipole is believed to be an example of such a Doppler-like effect, caused by the motion of our Local Group relative to the rest-frame of the smooth Hubble expansion. Similarly, the motion of an observer in section $\mathcal{A}$, relative to the mean bulk flow of that region, can induce a dipole-like anisotropy in the spatial distribution of the deceleration parameter. However, when the observer is a typical one -- with drift velocity close to the average flow of the patch -- the aforementioned dipolar anisotropy should be small. With these in mind we have considered anisotropic peculiar motions, analysed their kinematics and used the associated equations to discuss dipole-like anisotropies. Our results indicate that drift flows are generally consistent with some degree of dipolar anisotropy in the spatial distribution of $\tilde{q}$. Moreover, if such a $\tilde{q}$-dipole exists, the axis should not lie far from that of its CMB counterpart. This sounds plausible, given that both dipoles are induced by the drift motion of the Milky Way, though relative to different rest-frames in each case. Recently, there has been a number of reports suggesting that a small dipolar anisotropy might actually reside in the supernovae data~\cite{KL}. Moreover, the axis of the $\tilde{q}$-dipole appears to lie close to that of the CMB. In the literature, this has been largely treated as coincidental. Viewed from the perspective of this report, however, the alignment of the two dipolar axes seems more like a natural consequence rather than a mere coincidence.

\end{document}